# $Q$-Enhanced SH-SAW Ladder Filter in Thin-Film Lithium Tantalate Using Bartlett Apodization

Taran Anusorn, Tzu-Hsuan Hsu, Yuchen Ma, and Ruochen Lu
Chandra Department of Electrical and Computer Engineering, The University of Texas at Austin, Austin, USA
taran.anusorn@utexas.edu

***Summary*—Shear-horizontal surface acoustic wave (SH-SAW) filters have shown strong potential for low-loss, compact, GHz-frequency RF front ends. In this work, we demonstrate a high-performance SH-SAW filter design at 4.35 GHz utilizing 42°Y-cut thin-film lithium tantalate ($LiTaO_3$) on a $SiO_2$/Si platform. Despite the limitations of thin aluminum metallization and its associated ohmic losses, we show that implementing a Bartlett window apodization technique, primarily intended for in-band spurious-mode suppression, yields a significantly improved quality factor (*Q*) of 1,522 from 688 in conventional interdigitated SH-SAW resonators. This enhancement enables a third-order ladder filter at 4.3 GHz with an insertion loss of 1.59 dB, compared with 1.65 dB for a conventional SH-SAW filter. In addition, our filter with apodized resonator designs achieves a 3-dB fractional bandwidth (FBW) of 3.24% and out-of-band rejection exceeding 14 dB, all within a compact footprint of 0.4 mm². These results suggest that apodized thin-film $LiTaO_3$ designs are highly promising for low-loss, miniaturized, cost-effective radio-frequency acoustic solutions in next-generation communication and sensing applications.**

***Keywords—Acoustic filter; apodization; lithium tantalate; piezoelectric devices; shear horizontal surface acoustic wave.***

## I. Introduction

For radio-frequency (RF) and microwave applications below 6 GHz, surface acoustic wave (SAW) filters have remained the dominant technology due to their low insertion loss, compact footprint, and relatively simple fabrication process [1], [2]. Among them, shear-horizontal SAW (SH-SAW) devices implemented on piezoelectric-on-insulator (POI) platforms [3], [4], [5], [6], [7], particularly 42°Y-cut thin-film lithium tantalate ($LiTaO_3$) on silicon dioxide on silicon ($SiO_2$/Si) substrates [8], have attracted significant attention because of their high quality factor (*Q*), large electromechanical coupling ($k^2$), low temperature coefficient of frequency (TCF), and flexible frequency scaling. To suppress spurious modes and improve device performance, various dispersion-engineering techniques have been investigated, including material orientation optimization [9], tilted interdigitated transducers (IDTs) [10], piston-mode operation [11], slowness curve manipulation [12], miniaturized IDTs [13], crossed IDTs [14], and IDT apodization [15]. Despite these advances, the relatively low phase velocity of SH-SAW modes necessitates sub-micron IDT dimensions for operation above 4 GHz, posing substantial fabrication challenges and limiting achievable performance at higher frequencies.

To address this limitation, longitudinal leaky SAW (LL-SAW) devices are often employed in higher-frequency applications due to their significantly higher phase velocity, which enables improved frequency scaling, more relaxed fabrication requirements, and enhanced device performance [16], [17], [18]. However, LL-SAW devices typically rely on expensive substrates, such as silicon carbide (SiC) or diamond, to suppress acoustic leakage and maintain low loss. Consequently, there remains strong interests in extending SH-SAW technology into the C-band (4 - 8 GHz) using cost-effective substrate platforms. Motivated by this need, this work experimentally investigates the application of IDT apodization in SH-SAW resonators and filters operating above 4 GHz using sub-micron IDTs fabricated on a $LiTaO_3$/$SiO_2$/Si platform.

In this work, two third-order SH-SAW ladder filter prototypes employing conventional and Bartlett-window-apodized IDTs [Fig. 1] were fabricated together with their constituent standalone resonators. The proposed apodization significantly improves the *Q* of the constituent resonators, increasing the shunt resonator *Q* of 688 to 1,522 and the series resonator *Q* from 431 to 893. As a result, the apodized filter achieves a 1.59 dB insertion loss (IL) at 4.3 GHz with an improved 3 dB fractional bandwidth (FBW) of 3.24%, compared with 1.65 dB IL and 2.38% FBW for the filter employing conventional IDTs. Although the current IL remains limited by the thin IDT metallization, the demonstrated Q enhancement highlights the strong potential of apodized SH-SAW devices for realizing high-performance, cost-effective C-band acoustic filters.

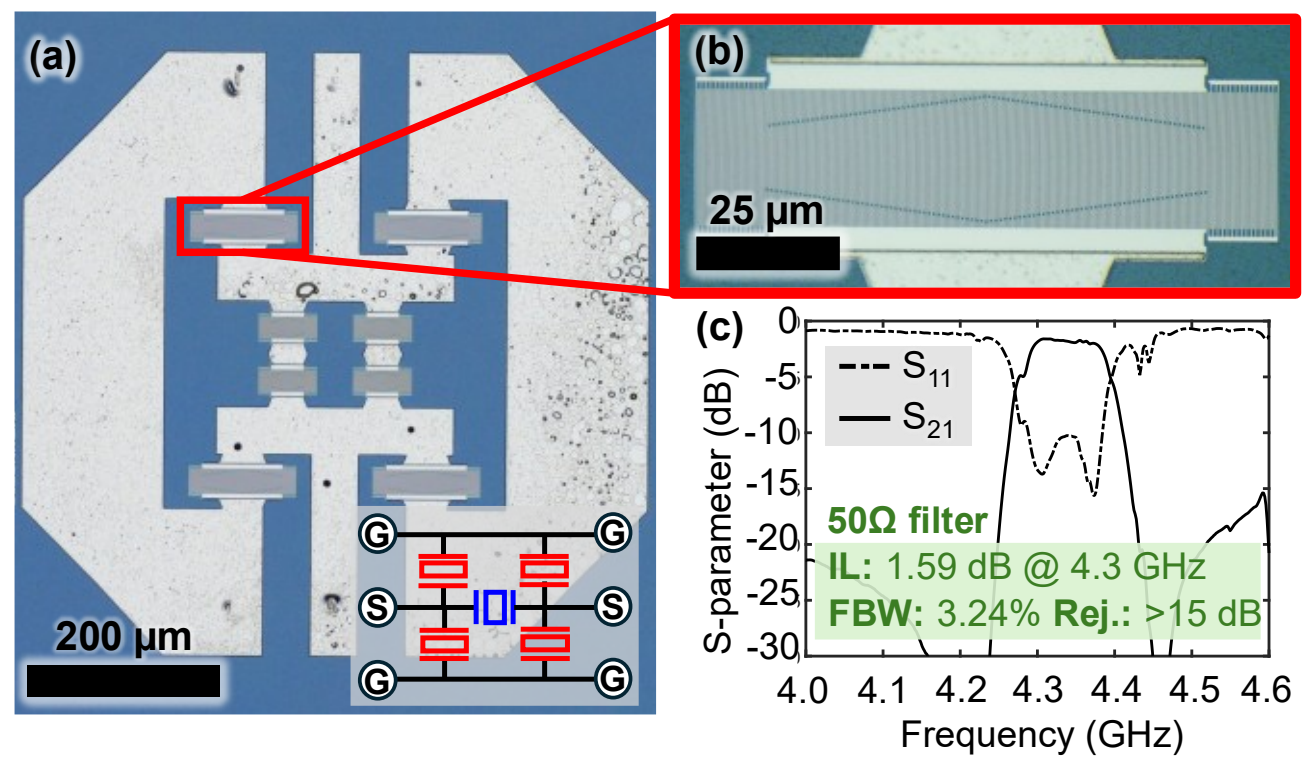

**Fig. 1.** (a) Fabricated third-order ladder filter with inset equivalent schematic diagram. (b) Zoomed-in optical image of the resonator, showing Bartlett-window apodized IDTs. (c) 50 Ω filter frequency responses.

## I. Design of Apodized SH-SAW Resonators and Filters

A generic SAW resonator is composed of IDTs for transduction between electrical RF signals and SAW

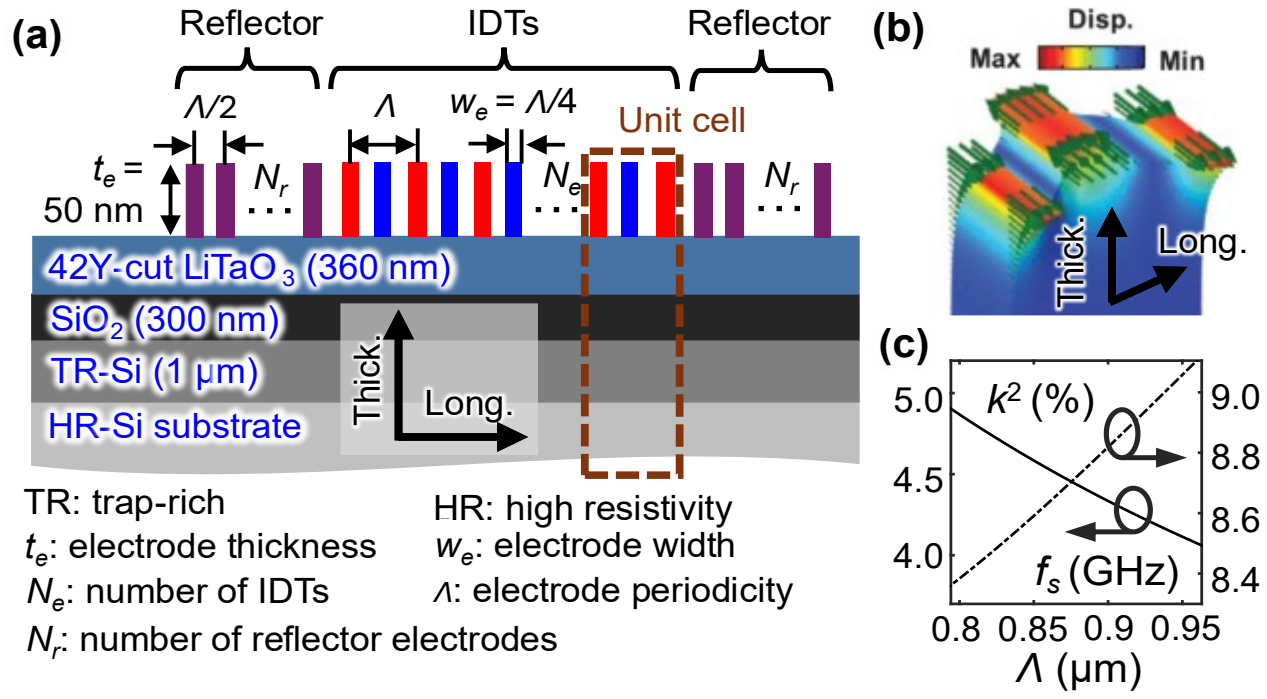

**Fig. 2.** (a) Cross-sectional schematic of implemented SAW resonators. (b) Simulated SH-SAW displacement mode shape. (c) Extracted $f_s$ and $k^2$.

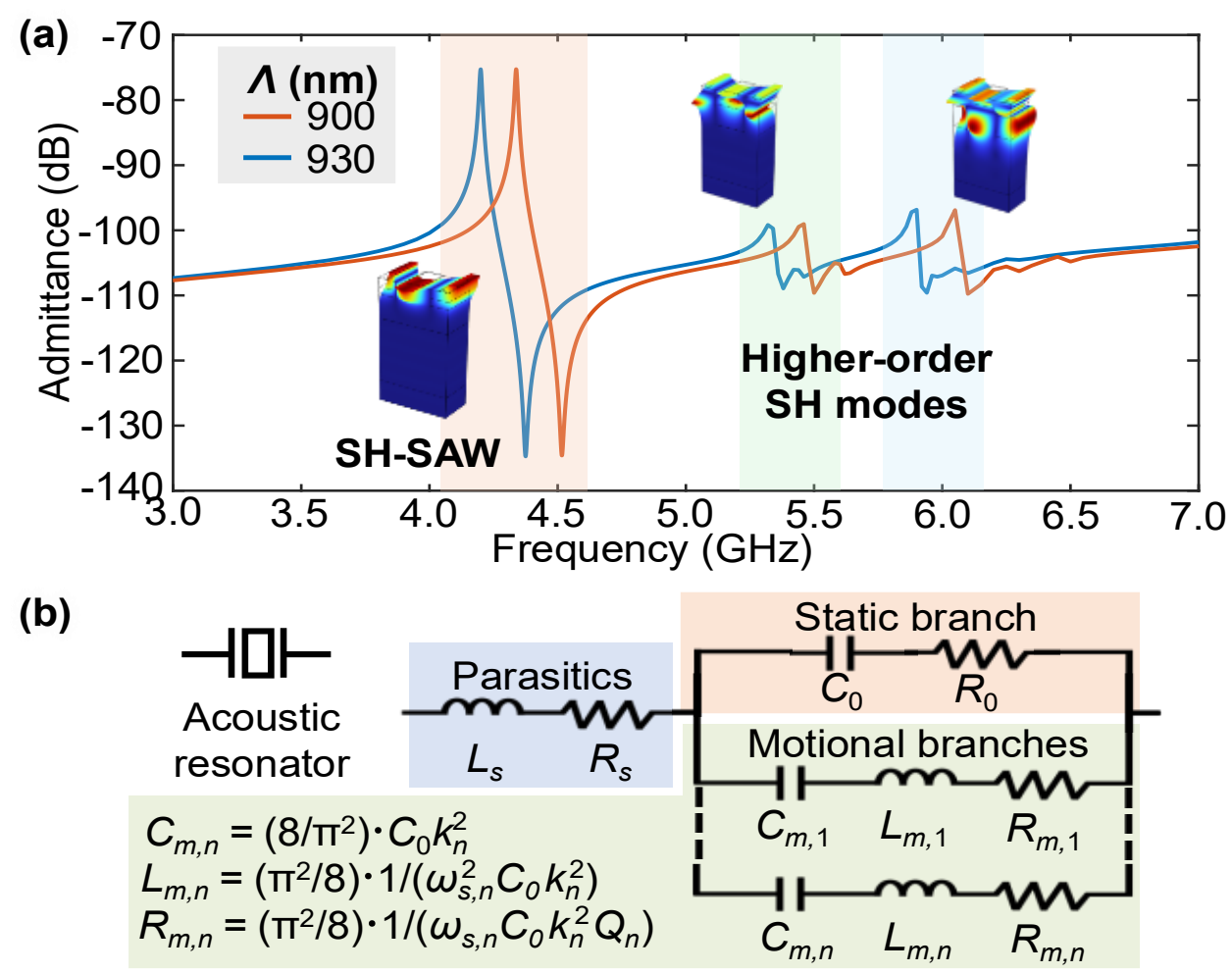

**Fig. 3.** (a) Simulated frequency responses of an SH-SAW resonator unit cell. (b) mBVD model with multiple motional branches.

propagating on the surface of a piezoelectric material, i.e., $LiTaO_3$ in this work. The acoustic waves are laterally confined by two shorted IDT arrays at both ends of the resonator, which serve as acoustic reflectors. Fig. 2(a) illustrates the cross-sectional structure of the SAW device together with the wafer-stack information provided by SOITEC. Notably, a $SiO_2$ layer facilitates temperature compensation while a trap-rich (TR) Si layer and high-resistivity (HR) Si substrate are incorporated to mitigate RF substrate losses.

To determine the frequency response of the SH-SAW resonator implemented on the provided wafer stack, an eigenfrequency analysis of a unit cell was conducted using COMSOL Multiphysics. The simulated SH-SAW mode shape is presented in Fig. 2(b). Fig. 2(c) shows the extracted dispersion curves obtained from a parametric study, illustrating the relationships of the series resonance frequency ($f_s$) and the perceived $k^2$ as functions of the IDT periodicity ($Λ$). The extracted results indicate that C-band operation near 4 GHz requires an IDT periodicity of approximately 900 nm.

Following the practical acoustic filter design methodology described in [19], the unit cell of the SH-SAW device was analyzed using frequency-domain finite-element simulations, and the resulting frequency responses, shown in Fig. 3(a), were subsequently fitted to a modified Butterworth-Van Dyke (mBVD) equivalent circuit model with multiple motional branches [20] to capture higher-order SH-SAW modes [Fig. 3(b)]. To demonstrate the design of C-band SH-SAW filters, a third-order ladder network based on the mBVD resonator model, beginning with a shunt resonator as shown in Fig. 3(a), was designed and optimized to achieve a 50 Ω-matched filtering response. Notably, only the branch representing the main tone was considered in the design for simplicity. The optimized filter response is presented in Fig. 3(b), together with the extracted mBVD parameters for each constituent resonator. Using the resonator static capacitance ($C_0$), the physical dimensions of each resonator, including the aperture length ($L$) and the number of IDT fingers ($N_e$), can be determined.

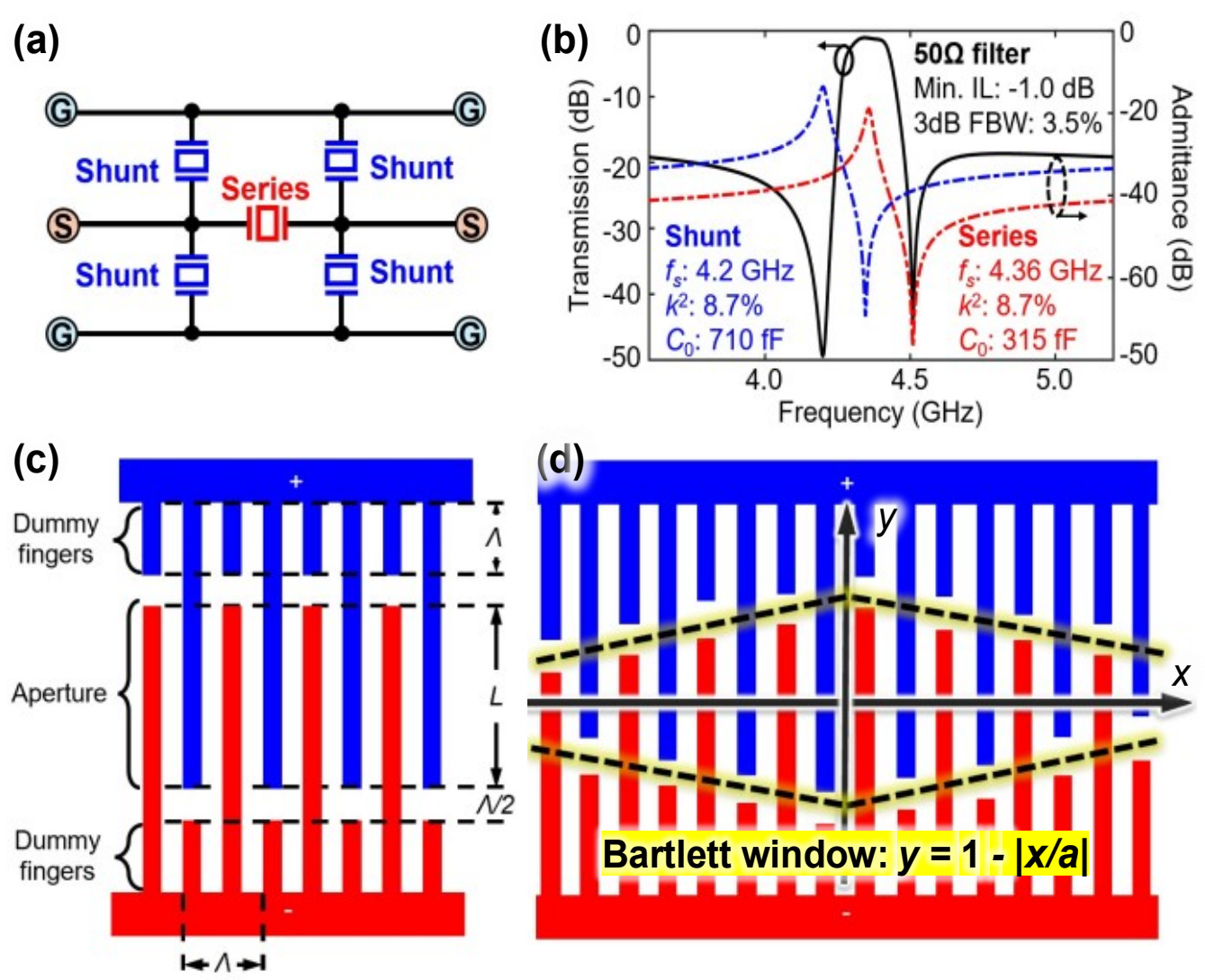

**Fig. 4.** (a) Third-order ladder filter with five resonators. (b) Simulated filter response with optimized mBVD design parameters. (c) Conventional and (d) Barlett-window apodized SAW IDT designs.

**TABLE I** DIMENSIONAL PARAMETERS FOR FILTER DESIGN

| | Series | | Shunt | |
|---|---|---|---|---|
| | Conventional | Apodized | Conventional | Apodized |
| $N_e$ | 39 | 92 | 29 | 69 |
| $Λ$ (μm) | 0.93 | 0.93 | 0.90 | 0.90 |
| $L$ (μm) | 24 | 24 | 21 | 21 |

For the conventional SAW resonator design shown in Fig. 4(c), dummy fingers were incorporated to reduce acoustic discontinuities caused by edge effects and to improve the fabrication yield of the sub-micron IDT structures. For the apodized design, only a Bartlett window function [Fig. 4(d)] was considered in this initial study, defined as

$$y = 1 - |x/a|, \qquad (1)$$

where $x$ denotes the position along the longitudinal direction and $y$ represents the normalized aperture profile in the transverse direction, with the center of the resonator defined as the origin. In this work, the Bartlett window parameter was chosen as $a$ = 0.5. For comparison purposes, $L$ was kept identical in both designs. Consequently, only $N_e$ was adjusted based on the effective electrode overlap area in the apodized structure to preserve the same $C_0$. Table I summarizes the key design parameters for both conventional and apodized resonators in the filter prototypes.

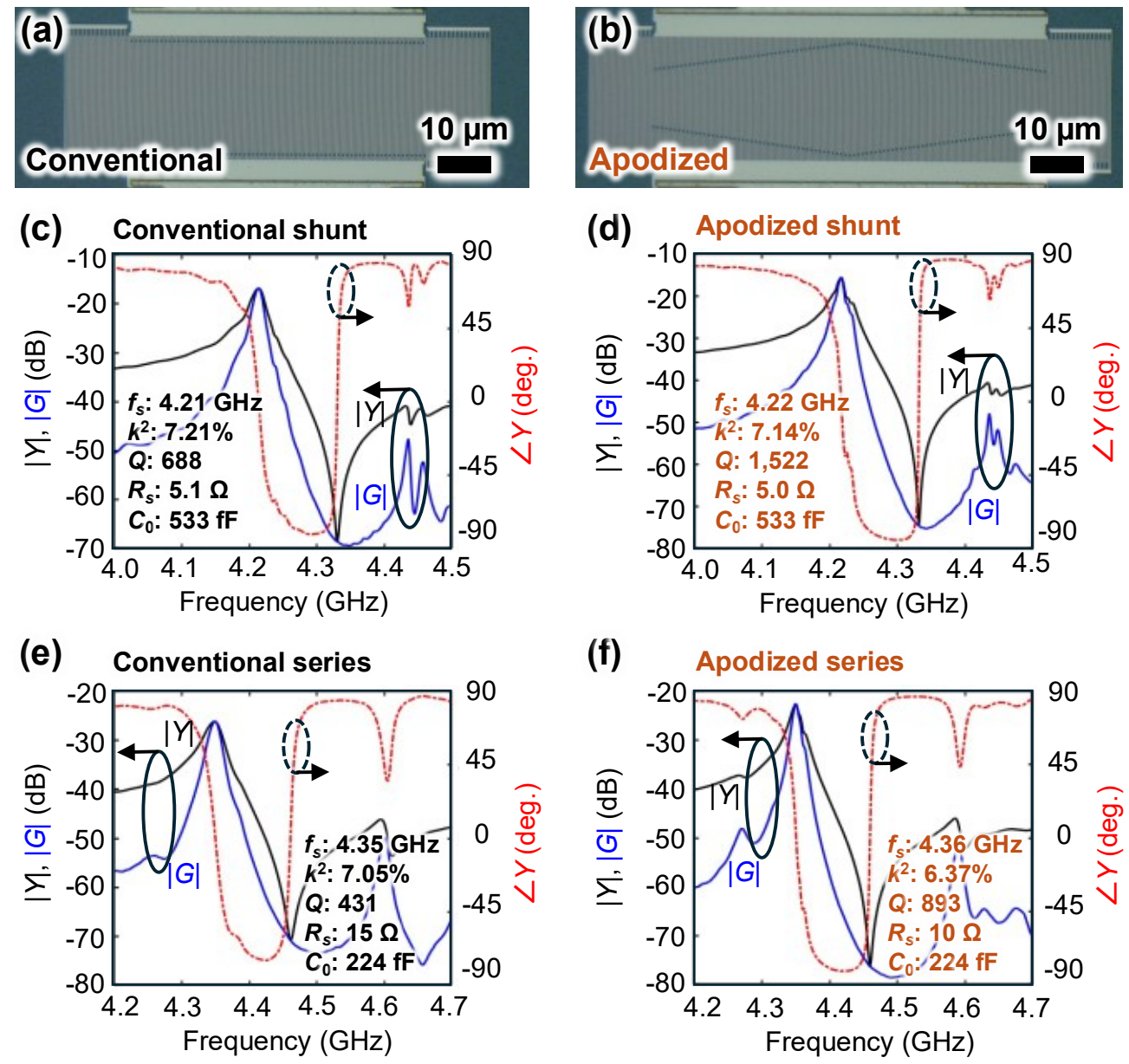

**Fig. 5.** Fabricated resonators with (a) conventional and (b) Bartlett-window apodized IDTs. Plots of admittance ($|Y|$: black) and conductance ($|G|$: blue) magnitudes and admittance ($\angle Y$: red) phase of shunt resonators with (c) conventional and (d) apodized IDTs and series resonators with (e) conventional and (f) apodized IDTs, along with their fitted mBVD parameters.

## II. Fabrication Results and Discussions

The filter prototypes and their standalone constituent resonators were fabricated by depositing aluminum (Al) via evaporation and lift-off techniques. The first deposition was 50 nm thick for electron beam lithographically defined IDTs and initial buslines. To reduce Ohmic losses due to the thin metal, an additional 550 nm Al deposition was performed to thicken the buslines of all devices.

Figs. 5(a) and 5(b) show optical images of the fabricated standalone conventional and apodized resonators, respectively. The measured admittances, together with their fitted mBVD parameters, of the fabricated resonators are compared in Figs. 5(c)-(f) for the conventional and apodized series and shunt resonators, respectively. Notably, the proposed Bartlett-window apodization improves the resonator $Q$ by more than a factor of two compared with the conventional design, despite a slight shift in resonance frequency and a modest reduction in $k^2$. Moreover, the fitted $C_0$ indicates that our IDT-based compensation scaling is accurate.

Interestingly, no in-band spurious modes are observed even in the conventional IDT design. This behavior is likely due to the relatively thin Al IDTs compared with the $LiTaO_3$ layer. As reported in [8], thicker Al IDTs combined with thinner $LiTaO_3$ films tend to concentrate more acoustic energy near the electrode region, thereby increasing the curvature of the slowness curve and exacerbating transverse spurious modes. However, the use of thinner IDTs introduces higher Ohmic losses, as reflected in the large series resistance ($R_s$) extracted from the fitted mBVD model.

Figs. 6(a) and 6(b) compare the Bode $Q$ of the fabricated conventional and apodized shunt resonators and the fabricated conventional and apodized series resonators, respectively. The Bode $Q$ was calculated using [21]

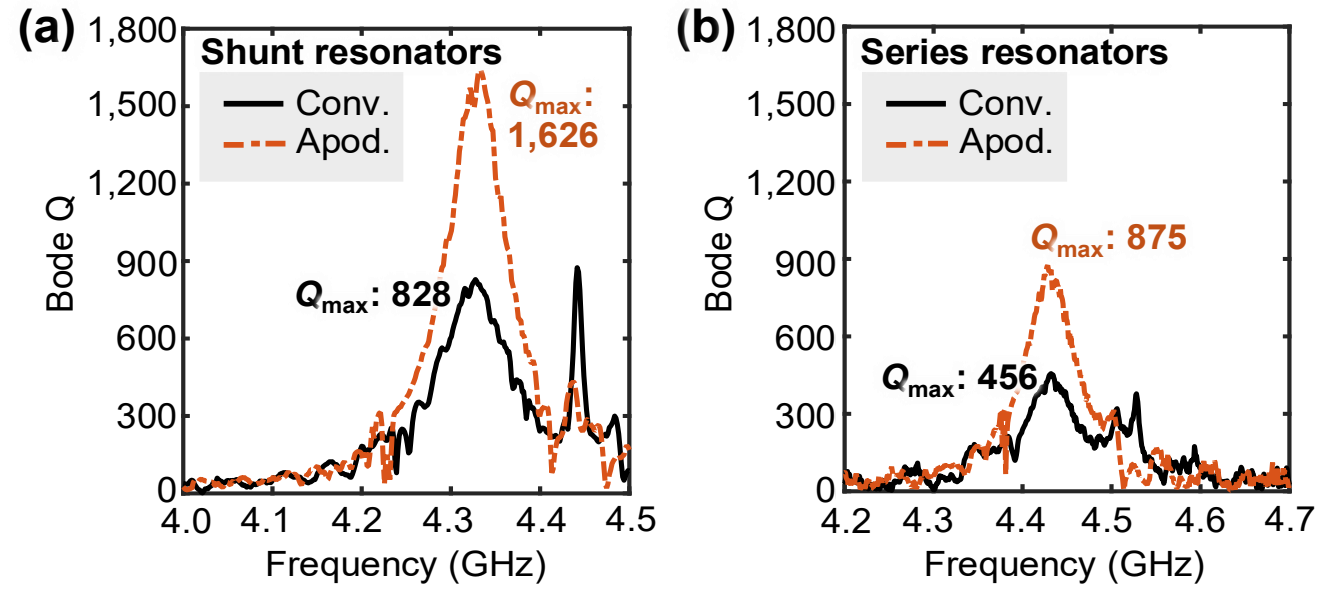

**Fig. 6.** Bode $Q$ of (a) shunt and (b) series resonators, for conventional (black) and Bartlett-window apodized (orange) IDTs.

$$Q_{\text{Bode}} = \omega \left| \frac{dS_{11}}{d\omega} \right| \frac{1}{1-|S_{11}|^2}. \tag{2}$$

It is evident that the apodized design outperforms the conventional design with the same trend observed in $Q$ from the mBVD model fitting. In addition to the improved resonator $Q$, the extracted Bode $Q$ responses of the apodized devices exhibit less spurious features compared with those of the conventional-IDT devices, although such spurious responses are not readily distinguishable from the admittance and conductance plots shown in Fig. 5. Furthermore, the apodization suppresses the $Q$ of the nearby Rayleigh-mode resonances excited around 4.45 GHz and 4.55 GHz in the fabricated shunt and series resonators, respectively.

In principle, IDT apodization can degrade the $Q$ of an SH-SAW resonator because of increased acoustic leakage at the apodized electrode boundaries. However, experimental studies have shown that this degradation becomes relatively minor when the resonator aperture is sufficiently large [22]. Based on the preliminary results obtained in this work, one possible explanation is that the larger resonant cavity of resonators with greater dimensions along the SH-SAW propagation direction contributes to higher $Q$ performance. This behavior is clearly reflected in Figs. 6(a) and 6(b), where the maximum Bode $Q$ of both the conventional and apodized shunt resonators is approximately twice that of their corresponding series resonators. It is therefore plausible that a conventional-IDT SH-SAW resonator with dimensions comparable to those of the apodized design could also achieve a higher $Q$. Nevertheless, the resonator dimensions and configurations must be carefully optimized to maintain the desired $C_0$ values, which are critical in the filter design. Consequently, further systematic investigations are required to fully understand the physical mechanisms underlying the proposed apodization and to establish comprehensive design guidelines for high-performance SH-SAW devices.

The measured S-parameters of fabricated 50 Ω-matched filters with conventional and Bartlett-window apodized IDTs are plotted in Fig. 7. The enhanced $Q$ of the apodized case not only reduces the minimum IL of the filter but also flattens the overall passband, which significantly improves the FBW from 2.38% to 3.24%. Nonetheless, the OoB rejection of both filters is limited by the nearby Rayleigh mode, which cannot be completely removed by apodization alone, as well as by other higher-order modes. This would require further studies of IDT optimization and dispersion engineering techniques [23]. Moreover, the main bottleneck is the high resistive loss $R_s$ of a series resonator, as evident in Table I, as well as that from the overall thin, long signal paths. This could be mitigated by thickening both IDTs and buslines, or using larger resonators in parallel, and by placing devices closer together. Moreover, in-depth studies on apodized IDTs could yield further insights into these new findings.

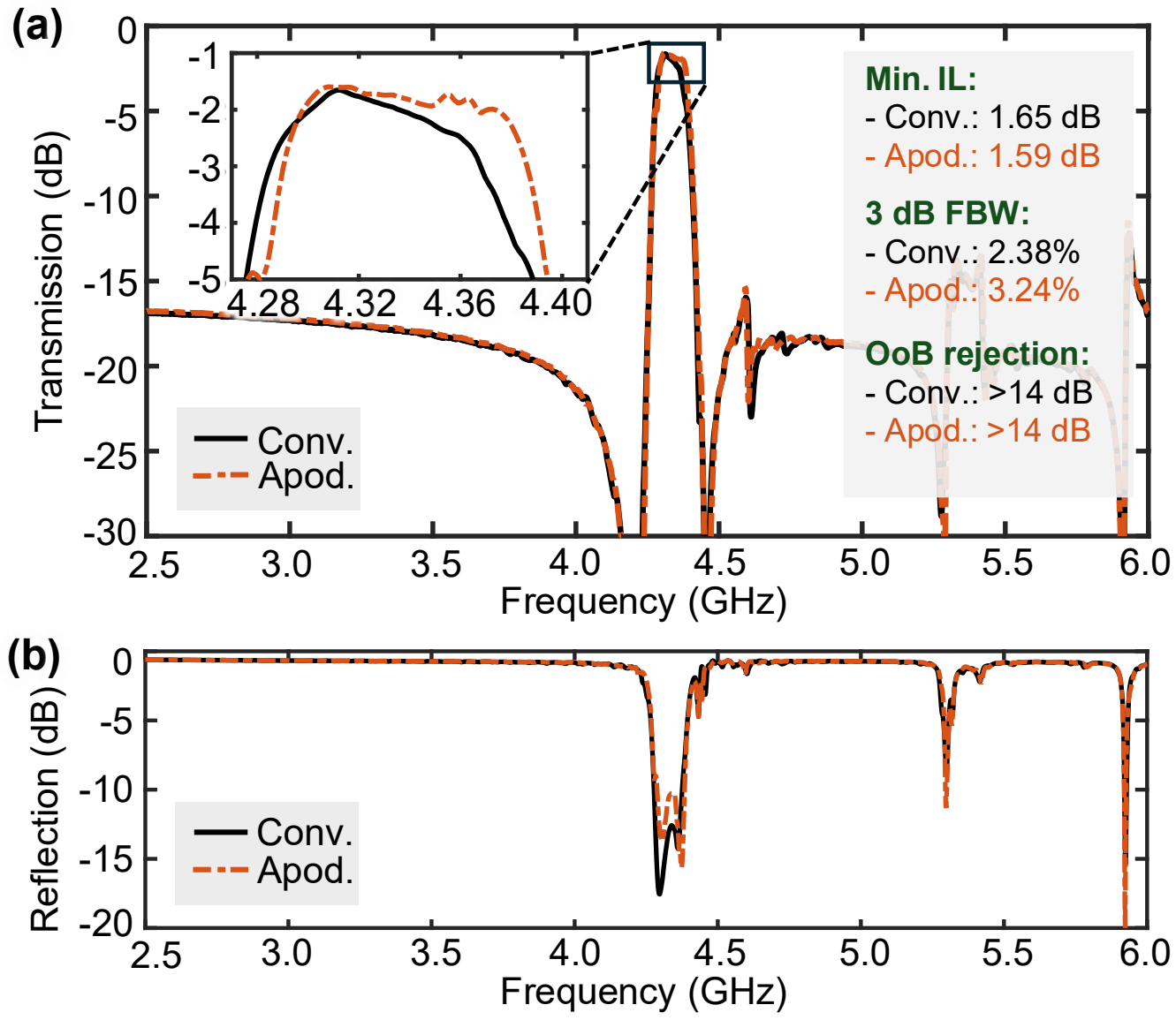

**Fig. 7.** (a) Transmission (with inset zoomed-in passband) and (b) reflection of filter prototypes with conventional (black) and apodized (orange) IDTs.

## III. Conclusion

This work presents a $Q$-enhanced SH-SAW filter with improved IL and a 3-dB FBW, employing Bartlett-window-apodized resonators on a cost-effective $LiTaO_3/SiO_2/Si$ POI platform. With further advancements, the proposed technique could offer a promising path toward compact, economical, high-performance RF filters.

## Acknowledgement

This work was supported by the National Science Foundation (NSF) under CAREER Award No. 2339731 and the Anandamahidol Foundation Scholarship. The author also thanks Dr. Harshvardhan Gupta, Mr. Jack Kramer, and Mr. Ian Anderson for their inspiring discussions.